\documentclass[onecolumn,nofootinbib,showpacs,prd,superscriptaddress, 10 pt]{revtex4}
\usepackage{amsmath}
\usepackage{amsfonts}
\usepackage{amssymb}
\usepackage{graphicx}
\usepackage{epstopdf}
\usepackage{subfigure}
\usepackage{epstopdf}

\begin{document}

\title{Different parametrizations of field equations  in the $f(R)$ theories of gravity}

\author{Liberato Pizza}
\affiliation{Dipartimento di Fisica, Universit\`a di Pisa, Largo Bruno Pontecorvo, i-56127, Pisa, Italy.}
\affiliation{Istituto Nazionale di Fisica Nucleare (INFN), Sezione di Pisa, Largo Bruno Pontecorvo, i-56127, Pisa, Italy.}
\begin{abstract}
In this paper we  describe two different  parametrizations of field equations in the $f(R)$ theories of gravity and  two  resulting  parametrizations of the Friedmann Equations. In particular we show how these two parametrizations lead to two different results for the  curvature pressure and  the curvature density obtained respectively as the $i-i$ component and the $0-0$ component of the curvature stress-energy tensor. We introduce formulas in order to pass from one parametrization to another one for the curvature pressure and the curvature density. Furthermore, we analyse under which conditions the two parametrizations of the curvature stress-energy tensor lead to significantly different results.
\end{abstract}


\maketitle
\section{$f(R)$ an overview}
Since the discovery of the current Universe speed up \cite{nobel}, many attempts to explain the cause of this phenomenon have been proposed. Among them, there are the $f(R)$ theories of gravitation \cite{formulations}.
The basic idea of  the $f(R)$ theories is to replace the standard Hilbert-Einstein Lagrangian density, i.e. $R$, with a generic function of $R$, i.e. $f(R)$.
Proceeding in this way the gravitational action becomes:
\begin{equation} \label{actionmetric}
S=\frac{1}{2\kappa} \int d^4x \,
\sqrt{-g}\, f(R)+S^{(m)} \;,
\end{equation}
where $\kappa$ is a coupling constant related to the gravitational constant $G$ and the speed  of light $c$, i.e. $\kappa\equiv\frac{8\pi G}{c^4}$, $g$ the spacetime involved and $S^{(m)}$ the action of the part of the matter.
In the last decade a lot of different functional forms of $f(R)$ have been proposed in order to explain  inflation, dark energy, dark matter and Bario-Leptogenesys \cite{bariolepto}. One  of the most studied model is the Starobinski one, i.e. $f(R)= R+ \alpha R^2$, that we will use in this paper \cite{rquadro}.
In order to determine the dynamical equations in $f(R)$ gravities, it is possible to vary Eq. (\ref{actionmetric}) with respect to the metric $g_{\mu\nu}$. This procedure leads to fourth order field equations of the form \cite{bariolepto, prima}:
\begin{equation}\label{metf}
f'(R)R_{\mu\nu}-\frac{1}{2}f(R)g_{\mu\nu}-\left[\nabla_\mu\nabla_\nu -g_{\mu\nu}\Box\right] f'(R)= \kappa
\,T_{\mu\nu}^{(m)}\,,
\end{equation}
where the prime, i.e. $"{'}"$, denotes the derivative with respect to $R$, and 
\begin{equation}\label{set}
T_{\mu\nu}^{(m)}\equiv\frac{-2}{\sqrt{-g}}\, \frac{\delta S^{(m)}
}{\delta g^{\mu\nu} }\,.
\end{equation}
 $\kappa$  will be imposed equal to 1 hereafter, and we will express explicitly the Einstein tensor $G_{\mu \nu}$ in 
 (\ref{metf}).
In the following  we will show two different parametrizations  of  the fields equations and of the resulting curvature stress-energy tensor. Furthermore, we will denote under which conditions the two parametrizations give significantly different values for the curvature pressure and density.
%%%%%%%%%%%%%%%%%%%%%%%%%%%%%%%%%%%%%%%%%%%%%%%%%%%%%%
\section{Parametrizations of field equations}
With the term of parametrization we want to denote the way to collect terms different from the Einstein Tensor, i.e. $G_{\mu \nu}=R_{\mu \nu} - \frac{1}{2}g_{\mu \nu} R$, in the left side of the fields equation (\ref{metf}). In principle we can organize them in a multitude of different ways, as will be clear hereafter.  
However, in this section, we present only two different schemes of parametrization for the field equations (\ref{metf}). These two schemes have been adopted in several articles, as can be seen when analysing the literature \cite{prima, seconda}. 
%%%%%%%%%%%%%%%%%%%%%%%%%%%%%%%%%%%%%%%%%%%%%%%%%%%%%%
\subsection{First Parametrization}
\noindent 
 The first parametrization,  often used in literature by  different authors as \cite{prima},  writes field equations like denoted in the following:
 \begin{equation} \label{field1}
G_{\mu\nu}= \frac{T_{\mu \nu}^{(m)}}{f'(R)} + T_{\mu \nu}^{curv}.
\end{equation}
In order to obtain this form for the field equations, we divide both sides of (\ref{metf}) for $f'(R)$ and  we obtain:
\begin{equation}\label{me}
R_{\mu\nu}-\frac{1}{2}\frac{f(R)}{f'(R)}g_{\mu\nu}-\frac{1}{f'(R)}\left[\nabla_\mu\nabla_\nu -g_{\mu\nu}\Box\right] f'(R)= 
\,\frac{T_{\mu\nu}^{(m)}}{f'(R)}\,,
\end{equation}
then adding and subtracting  $-\frac{1}{2} g_{\mu \nu} R$ to the left hand side of the previous equation, we obtain:
\begin{equation}\label{m}
G_{\mu\nu}+ \frac{1}{2} g_{\mu\nu} R -\frac{1}{2}\frac{f(R)}{f'(R)}g_{\mu\nu}-\frac{1}{f'(R)}\left[\nabla_\mu\nabla_\nu -g_{\mu\nu}\Box\right] f'(R)= 
\,\frac{T_{\mu\nu}^{(m)}}{f'(R)}\,.
\end{equation}
The curvature stress-energy tensor, responsible for curvature corrections, becomes:
\begin{eqnarray}
T^{curv}_{\mu \nu} & = &- \Big( \frac{1}{2} g_{\mu \nu} R- \frac{1}{2}\frac{f(R)}{f'(R)}g_{\mu\nu}- \frac{1}{f'(R)} (\nabla_{\mu} \nabla_{\nu} f'(R)- g_{\mu \nu} \square f'(R))\Big). \label{eq:curvstress}
\end{eqnarray}
Note that the semicolon, i.e. $"^{;}"$, stands for the covariant derivative. The $T^{curv}_{\mu \nu}$ in (\ref{eq:curvstress}) represents a source for the entire stress-energy tensor and as it has been done in literature it permits to describe the Universe dynamics  \cite{bariolepto, prima}. 
 \\ Obviously,  looking at Eq. (\ref{field1}), the matter tensor is coupled to the Ricci scalar by means of ${T}^{(m)}_{\mu \nu}/f'(R)$. Hereafter we consider a standard Friedmann-Lemaitre-Robertson-Walker metric (FLRW), i.e.  $ds^2= dt^2-a(t)^2\left(dr^2+r^2 d\Omega^2\right)$ hence the corresponding modified Friedmann equations obtained reading the $0-0$ and $i-i$ components of the field equations are \cite{bariolepto}:
\begin{equation}\label{eq: fried1}
H^2  = \frac{1}{3} \left [ \rho_{curv} +
\frac{\rho_m}{f'(R)} \right ]\,,
\end{equation}
and
\begin{equation}\label{eq: fried2}
2\dot{H} +3 H^2 = - P_{curv}- \frac{P_{m}}{f'(R)}\,,
\end{equation}
where the dot denotes the derivative with respect to cosmic time, and  $\rho_m$  and $P_m$ are respectively the matter density, i.e. the $0-0$ component of $T_{\mu \nu}^{(m)}$, and the matter pressure, i.e. the $i-i$ component of $T_{\mu \nu}^{(m)}$.
In these equations we have posed  equal to $0$ the scalar curvature, i.e.  $k=0$, in  agreement with recent observations  and we define the curvature density as \cite{bariolepto, prima}:
\begin{equation}\label{eq: rhocurv}
\rho_{curv} =   \frac{1}{2} \left( \frac{f(R)}{f'(R)}  - R \right) - 3 H \dot{R} \frac{f''(R)}{f'(R)} \,,
\end{equation}
 the curvature pressure as
\begin{equation}
P_{curv} = \omega_{curv} \rho_{curv} \label{eq: pcurv}\,,
\end{equation}
where $\omega_{curv}$  is straightforwardly  given by
\begin{equation}
\omega_{curv} = -1 + \frac{\ddot{R} f''(R) + \dot{R} \left [ \dot{R}
f'''(R) - H f''(R) \right ]} {\left [ f(R) - R f'(R) \right ]/2 - 3
H \dot{R} f''(R)}\,, \label{eq: wcurv}
\end{equation}
and it represents the  equation of state (EoS) parameter for the curvature part \cite{formulations}.\footnote{Useful relationships to obtain the analytical expression of the curvature density and pressure are collected in the Appendix A.}
  The relation between the  Ricci scalar and the Hubble rate, that will be used in the following, is 
\begin{equation}
R = -6 \left ( \dot{H} + 2 H^2\right )\,.
\label{eq: constr}
\end{equation}

\subsection{Second parametrization}
In the second parametrization of Eq. (\ref{metf}), we rewrite, in order to be more explicit, the field equations (\ref{metf}):
\begin{equation}\label{metf2}
f'(R)R_{\mu\nu}-\frac{1}{2}f(R)g_{\mu\nu}-\left[\nabla_\mu\nabla_\nu -g_{\mu\nu}\Box\right] f'(R)= 
\,T_{\mu\nu}^{(m)}\,,
\end{equation}
and, instead of dividing both members for $f'(R)$, we, at first, add and subtract the Einstein tensor, $G_{\mu \nu}$ i.e. $R_{\mu \nu}- \frac{1}{2}g_{\mu \nu} R$  so we get:
\begin{equation}
f'(R)R_{\mu\nu}-\frac{1}{2}f(R)g_{\mu\nu}-\left[\nabla_\mu\nabla_\nu -g_{\mu\nu}\Box\right] f'(R) +R_{\mu \nu}- \frac{1}{2}g_{\mu \nu} R -R_{\mu \nu}+ \frac{1}{2}g_{\mu \nu} R = 
\,T_{\mu\nu}^{(m)}\,.
\end{equation}
Thus we can rewrite field equations like:
\begin{equation}
G_{\mu \nu}-   T_{\mu \nu}^c=  T_{\mu \nu}^{(m)},
\label{field}
\end{equation}
where $T_{\mu \nu}^c$, i.e. the  curvature stress-energy tensor,  in this case becomes:

\begin{equation}
 T_{\mu \nu}^c= (1- f'(R)) R_{\mu \nu}+ \frac{1}{2}g_{\mu \nu} (f(R) -R)+\left[\nabla_\mu\nabla_\nu -g_{\mu\nu}\Box\right] f'(R).
\end{equation}
Proceeding as done so far,  in this case we have obtained another parametization for the curvature stress-energy tensor, now denoted as $T_{\mu \nu}^c$, and consequentially new different expressions for the curvature pressure and density are:
\begin{equation}
\rho^c=-3(1-f'(R))(\dot{H}+ H^2)  +\frac{1}{2}(f(R)-R) - 3 H f''(R) \dot R,
\end{equation}
\begin{equation}
P_c=(1-f'(R))(\dot{H}+ 3 H^2)- \frac{1}{2}(f(R)-R)+f'''(R) \dot{R}^2+f''(R) \ddot{R}+2 H f'' \dot{R}.
\end{equation}
If we take $0-0$ component of (\ref{field}) we get the first Friedmann equation:
\begin{equation}
H^2=\frac{1}{3}(\rho_m +\rho_c) \label{fr 1}
\end{equation}
and for the $i-i$ component we get:
\begin{equation}
2 \dot{H} + 3 H^2=-P_m- P_c.\label{fr 2}
\end{equation}
This parametrization is, for example, used  in \cite{seconda}.
\section{The deviation parameter}
In this section we introduce a new parameter useful to quantify the difference between the curvature stress-energy tensors in the two parametrizations. 
Both sets of Friedmann equations, i.e. (\ref{eq: fried1}, \ref{eq: fried2}) and (\ref{fr 1}, \ref{fr 2}), must obviously have the same solution for the Hubble-rate, because they have been obtained starting from the same action and  field equations (\ref{actionmetric}), (\ref{metf}). 
In the following we highlight   different  results for the components of the curvature tensor in the two different parametrizations, which from here onwards we call, for simplicity, system A and system B, in order of apparition in this paper.

At first, we highlight that in A the matter density is divided by $f'(R)$, i.e. $\rho_{eff}= \frac{\rho_m}{f'(R)}$, thus a modification of the gravity, i.e. $f(R)$ in the A system, generates an effective matter density that is reduced with respect to the real one. The same condition is realized for the pressure: $P_{eff}=  P_m/ f'(R)$. However the parameter of state for matter, i.e $\omega_m=P_m/\rho_m$ does not change, for the simplification of $f'(R)$.
Because  the  Hubble rate computed in A and B must be the same, we can find the relationship between $\rho_c$ and $\rho_{curv}$, $P_c$ and $P_{curv}$.
Equating (\ref{eq: fried1}) and (\ref{fr 1}), (\ref{eq: fried2}) and (\ref{fr 2}) a simple computation gives:
\begin{equation}\label{D}
\rho_c- \rho_{curv}= \rho_m  \Big(\frac{1}{f'(R)} -1 \Big),
\end{equation}
\begin{equation}
P_c- P_{curv}=P_m \Big(\frac{1}{f'(R)} -1\Big).
\end{equation}
It is worthwhile to note that the second sides of both equations denote the difference between the effective matter density (pressure) in the system A and B. Indeed  $P_m$  is the pressure of the component of the Universe in B, while $P_m/f'(R)$ is the effective pressure in A. 
Now we calculate the relationship between  $\omega_c$ and $\omega_{curv}$ in the two systems. 
In order to perform this computation we rewrite  previous relationships as:
\begin{equation}
\rho_c= \rho_{curv} + \rho_m \Big(\frac{1}{f'(R)}-1\Big)
\end{equation}
\begin{equation}
P_c=P_{curv}+P_m \Big(\frac{1}{f'(R)}-1 \Big).
\end{equation}
Furthermore the parameter of state for dark fluid can be denoted as:
\begin{equation} 
\omega_c= \frac{P_c}{\rho_c}= \frac{P_{curv}+ P_m D}{\rho_{curv}+ \rho_m D},
\end{equation}
with $D= \frac{1}{f'(R)}-1$, which we name the Deviation parameter.
The previous one becomes:
\begin{equation} \label{2}
\omega_c= \frac{\rho_{curv} \omega_{curv}+\rho_m D \omega_m}{P_{curv} (\frac{1}{\omega_{curv}})+ P_m D (\frac{1}{\omega_m})}.
\end{equation}
We have baptized $D$ the deviation parameter for his behaviour: for $D\rightarrow 0$ we obtain $\omega_c=\omega_{curv}$, $ P_c=P_{curv}$ and $\rho_c= \rho_{curv}$. $D$ is equal to zero if $f(R)=R$, i.e. the extended theory of gravity reduces to General Relativity. $D$ is slightly different from zero if the $f(R)$ analysed can be written as $f(R) \approx R+ \alpha R^2 + \gamma R^3 +....$, with all non linear terms in $R$   small perturbations and sub dominant respect to the linear one, at a given cosmic time.
If  $D$ becomes big the components of two  curvature stress-energy tensors, i.e. $T^{\mu \nu}_{curv}$ and $T^{\mu \nu}_{c}$, present a wider difference.

In order to be clearer we are going to exhibit some examples. Thus adopting the Hubble rate solution $H\approx \frac{\beta}{t}$ (with $t$ cosmic time), we itemize, in the  following table, the different analytical values of $\omega_c$ and $\omega_{curv}$, computed for some $f(R)$ models, i.e.

\begin{figure}
\includegraphics[scale=1]{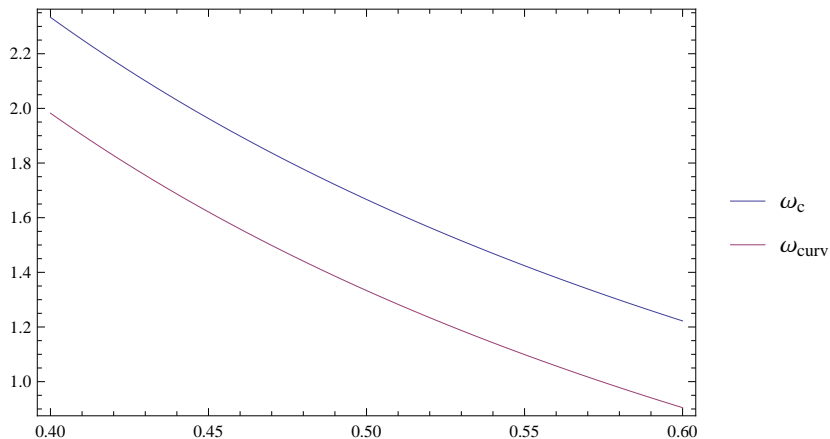}
\caption{Here the behaviour of $\omega_c$, $\omega_{curv}$ computed for the model $f(R)=R+ \alpha R^2$, for the approximated solution $H\approx \beta/t$ in the range of $\beta \in [0.4,0.6]$.}   \label{fig2}
\end{figure}

\begin{center}
    \begin{tabular}{ | l | l | l | p{5cm} |}
    \hline
    $f(R)$ &$\omega_c$ & $\omega_{curv}$  \\ \hline
    $R+\alpha R^n$ & $\frac{2 n}{3 \beta }-1$ &$\frac{6 \beta ^2-3 \beta -4 n^2+4 \beta  n+2 n}{-6 \beta ^2+3 \beta -6 \beta  n} $ \\ \hline
    $R+ \alpha R^2 $ & $\frac{4}{3 \beta }-1$ & $\frac{-6 \beta ^2-5 \beta +12}{6 \beta ^2+9 \beta } $ \\ \hline
    $R^2$ & $\frac{18 \left(6 \beta ^2-11 \beta +4\right)+(2-3 \beta ) t^2}{3 \beta  \left(-36 \beta +t^2+18\right)} $ & $ \frac{-6 \beta ^2-5 \beta +12}{6 \beta ^2+9 \beta }$  \\
    \hline
   $R+ \alpha R^2 + \gamma R^3$ & $\frac{\alpha  (3 \beta -4) t^2-6 \beta  \left(2 \beta ^3+15 \beta ^2-48 \beta +20\right) \gamma }{3 \beta  \left(2 \beta  \left(2 \beta ^2+19 \beta -10\right) \gamma -\alpha  t^2\right)} $ & $ \frac{\alpha  \left(6 \beta ^2+5 \beta -12\right) t^2-36 \beta  \left(4 \beta ^3+4 \beta ^2-23 \beta +10\right) \gamma }{3 \beta  \left(12 \beta  \left(4 \beta ^2+8 \beta -5\right) \gamma -\alpha  (2 \beta +3) t^2\right)} $ \\ \hline
\end{tabular}.
\end{center}
In particular in  Figure \ref{fig2} we  graph the behaviour of $\omega_c$ and $\omega_{curv}$ for the widely studied $f(R)=R+\alpha R^2$. 
Furthermore we introduce another meaning of the Deviation Parameter. From eq. (\ref{D}) we can see that the Deviation parameter can be denoted as:
\begin{equation}
D=\frac{\rho_c -\rho_{curv}}{\rho_m},
\end{equation}
and in this way $D$ quantifies how much the two different curvature densities in A and B differ from the background  density (e.g. matter density, radiation density).  The same definition can be extended to the pressure. Again, the $D$ parameter can be used  to quantify how  much A and B system give different results for  curvature components, respect to the background.

As highlighted before, for $f(R)= R$ or for small polynomial perturbations from Einstein-Hilbert action, $f'(R)\approx 1$ and $D\approx 0$.
However this condition is dependent by the time at which we are studying Universe Dynamics. In this way, given a particular $f(R)$ theory, chosen to describe gravitational force in the Universe, can exist a time at which the non-linear terms become dominant and the $D$ parameter increase. Consequently the two parametrizations for the curvature stress-energy tensors become widely different. 

\begin{figure}[htbp]
\centering%
\subfigure[{ The $3D$ plot of $D(R, \alpha)$}\label{11}]%
{\includegraphics[scale=0.5]{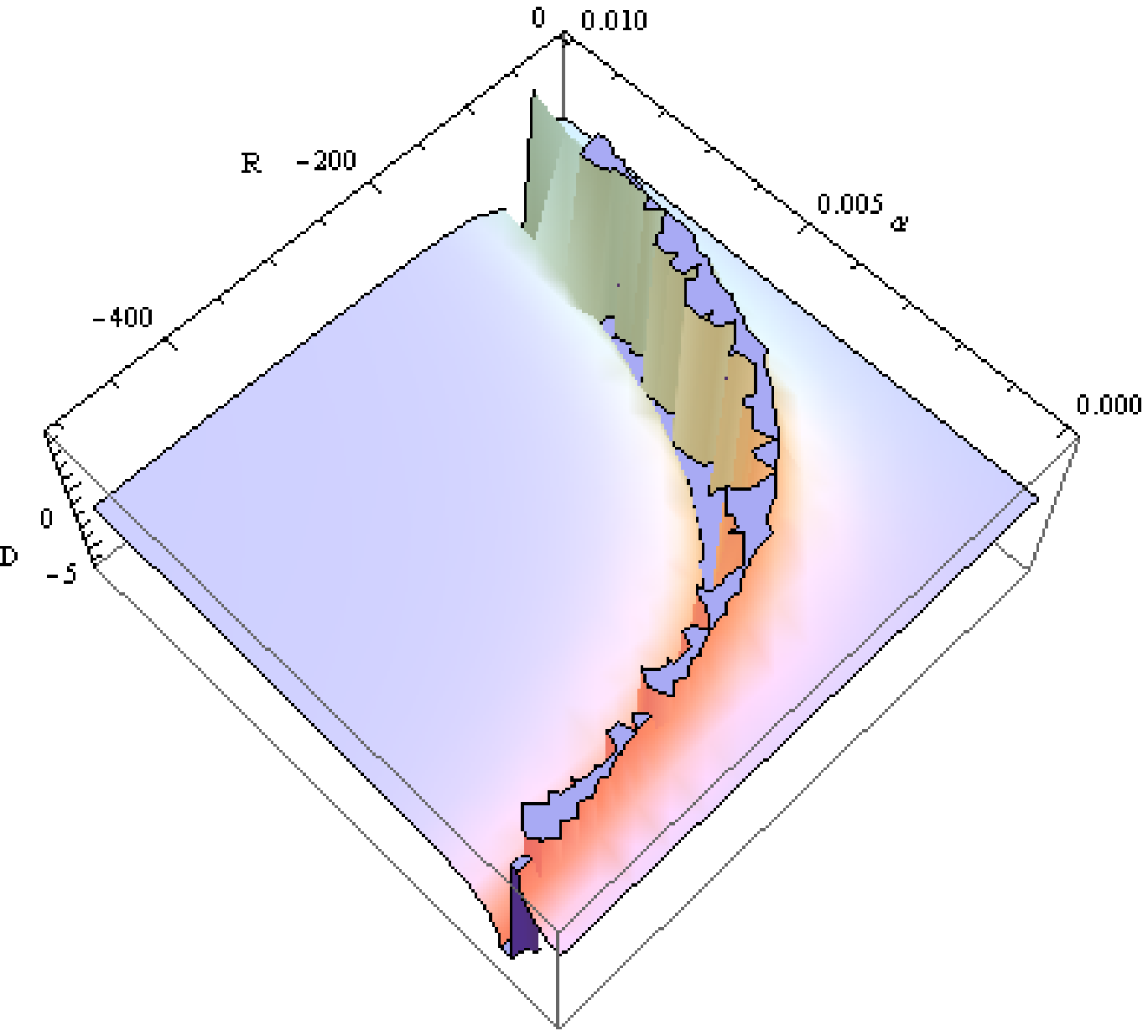}}\qquad\qquad
\subfigure[{The Contour plot of $D(R, \alpha)$}\label{12}]%
{\includegraphics[scale=0.5]{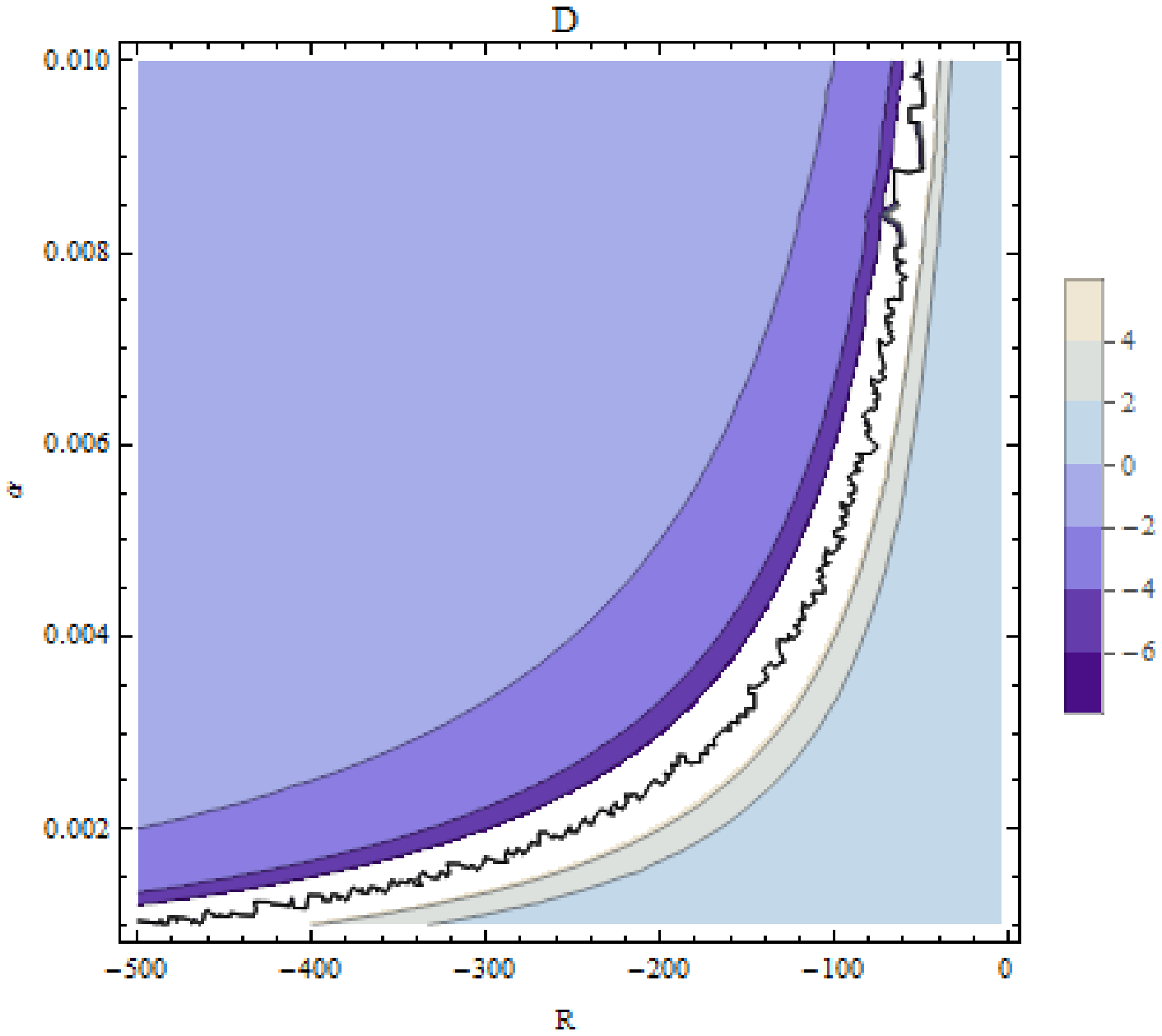}}\qquad\qquad
\caption{In this Figure we show the behaviour of $D$ for different values of $R$ and $\alpha$ in the $R+\alpha R^2$ model.  \label{1}}
\end{figure}

In order to better understand what we said we study the case of one of the most studied $f(R)$, i.e  $R+ \alpha R^2$. This model could be a valid phenomenological candidate to explain dark energy, inflation, and other early Universe phenomenon like Leptogenesis \cite{bariolepto}.
As known by Friedmann equation (\ref{eq: fried1}) the Ricci scalar is a function of the cosmic time or equivalently of the redshift parameter $z$. So if we consider $f'(R)= 1+ 2\alpha R$ there are  values of $R$ (and consequent values of cosmic time)  for which the $D$ function tends to be big. Thus  $(R,\alpha)$ points in proximity of the hyperbole $ 1+ 2\alpha R=\epsilon$, with $\epsilon$ small, have a large $D$, as shown in  Figure \ref{1}.
In proximity of these values of $(R,\alpha)$, components of two curvature stress-energy tensor give results widely different. Consequently the effective interpretation of  extra curvature components as a source tensor could be misleading.

\section{Conclusions and perspectives}
In this paper we have described two parametrizations of the curvature stress-energy tensor which emerge denoting the extra geometrical terms of curvature in field equations as a stress-energy tensor. These two parametrizations of the tensor have been denoted as $T_{\mu \nu}^c$ and $T_{\mu \nu}^{curv}$, and similarly it has been done for its respective component, i.e. pressure and density.
We have described how it is possible to transform  from one parametrization to another one (and vice-versa) by means of  the introduction of the Deviation parameter. The Deviation parameter takes into account the emerging difference between the components of the two curvature stress-energy tensor. Studying  the behaviour of the deviation parameter we have argued under which conditions the two parametrizations for the tensor (also denoted with  A and B) give significantly different results for the curvature pressure and density. In particular we note that for small perturbation from the General Relativity action  the results of A and B systems are very similar and $D\approx 0$. While if the non linear terms of the action dominate, at a given cosmic time, the curvature stress-energy tensor give hugely differents results for curvature pressure and curvature density in A and B. In this case to denote the extra curvature components present in the field equations (\ref{metf}) as a stress-energy tensor can be misleading, being the results for curvature pressure, curvature density and $EoS$ parameter depending from the parametrization chosen. In this case it could be better to treat the extra components as geometrical terms, which add contributes to the Einstein tensor, i.e. $G_{\mu\nu} + G_{\mu\nu}^c$.
Certainly these two parametrizations, described here, are only two example of a wider class of parametrizations that can be achieved. Further studies about the different descriptions of the dark matter or dark energy phenomenon by means of curvature stress-energy tensor should be done.

\section{Appendix: Useful Relations}
In this Appendix we report some useful relations that we use to obtain the pressure and density from (\ref{metf}), i.e. field equations \cite{seconda}.

The Ricci $0-0$ component is given by:
\begin{equation}
R^0_0= -6(\dot{H}+ 2H^2),
\end{equation}
while the $i-i$ is
\begin{equation}
R^i_i=- (\dot{H} + 3 H^2).
\end{equation}
The D'Alambertian opertor applied to the $f(R)$ can be rewritten as:
\begin{equation}
\square f'(R)=\ddot{f}'(R)+3 H\dot{f}' (R),
\end{equation}
considering that the time and spatial component can be denoted as:
\begin{equation}
\nabla_0\nabla_0 f'(R)= \ddot{f}'(R)
\end{equation}
\begin{equation}
\nabla_i\nabla^jf'(R)= H \dot{f}'(R) \delta_i^j, \text{ where}
\end{equation}
\begin{equation}
\dot{f}'(R)=f''(R) \dot{R}, \quad \ddot{f}'(R)= f'''(R) \dot{R} ^2 + f''(R)\ddot{R}.
\end{equation}

The continuity equation is valid both for  $T_{\mu\nu}^c$ component and  and for fluid $m$. For each of them can be written the continuity equation:
\begin{equation}
\dot{\rho}+ 3 H(\rho+ P)=0,
\end{equation}
with solution
\begin{equation}
\rho= \rho_0 e ^{-3 \int H (1+\sigma) dt},
\end{equation}
with $\rho_0$ constant and $\sigma= P/\rho$.

For  constant $\sigma$, in the Universe phase considered, we obtain the following result for density:
\begin{equation}
\rho=\rho_0 a^{-3(1+\sigma)}.
\end{equation}
\section*{Acknowledgements}
%%%%%%%%%%%%%%%%%%%%%%%%%%%%%%%%%%%%%%%%%%%%%%%%%%%%%%

 The author wishes to express his gratitude to G. Lambiase and V.Galluzzi for discussions on the topic of present work and A. Strumia for his support. The author is funded by INFN section of Pisa and University of Pisa.

%%%%%%%%%%%%%%%%%%%%%%%%%%%%%%%%%%%%%%%%%%%%%%%%%%%%%%


\begin{thebibliography}{99}
%%%%%%%%%%%%%%%%%%%%%%%%%%%%%%%%%%%%%%%%%%%%%%%%%%%%%%

\bibitem{nobel}
A. G. Riess, et al., Astron. J, {\bf 116}, 1009, (1998); S. Perlmutter, et al., ApJ, {\bf 517}, 565, (1999).

\bibitem{formulations}
A. De Felice, S. Tsujikawa, Living Rev. Relativity {\bf 13}, 3, (2010);
T. P. Sotiriou, V. Faraoni, Rev. Mod. Phys. {\bf 82}, 451 ,(2010); S. Capozziello, V. Faraoni, \emph {Beyond Einstein gravity}, Springer, (2011); S. Capozziello, M. De Laurentis, V. Faraoni, The Open Astronomy Journal, {\bf 3}, (2010), ArXiv:0909.4672v2[gr-qc], (2009).



\bibitem{bariolepto}  S. Nojiri, N. Odintsov,  arXiv:1408.3561v3 [hep-th] (2014); A. A. Starobinsky, JETP Lett.86 157, 2007; G. Lambiase, S. Mohanty, L. Pizza,  Gen. Relativ. and Gravit., {\bf 45}, 1771, (2013); L. Pizza,  arXiv:1411.5348v2 (2014); M. Artymowski, Z. Lalak,  arXiv:1405.7818v2 (2014); M. Rinaldi, G. Cognola, L. Vanzo, S. Zerbini,  arXiv:1410.0631v4, (2015); Y. S. Myung, arXiv:1503.03559v1, (2015); L.Pizza, arXiv:1506.08321v2 (2015); S. Capozziello, V. Galluzzi, G.Lambiase, L.Pizza, arXiv:1507.06835 (2015), S. Capozziello, V.F. Cardone, and A. Troisi, JCAP0608:001, (2006).



\bibitem{rquadro}
A. A. Starobinsky, Phys. Lett. B 91, 99 (1980).

\bibitem{prima} S. Capozziello, Int. J. Mod. Phys. D, {\bf 11}, 483, (2002).S. Capozziello, M. De Laurentis, O. Luongo, Annal. Phys., {\bf 526}, 309-317, (2014); A. Aviles, A. Bravetti, S. Capozziello, O. Luongo, Phys. Rev. D, {\bf 90}, 043531, (2014); A. Aviles, A. Bravetti, S. Capozziello, O. Luongo, Phys. Rev. D, {\bf 90}, 043531, (2014); C. Gruber, O. Luongo, Phys. Rev. D, {\bf 89}, 103506, (2014);


\bibitem{seconda}
 S. Capozziello, M. De Laurentis, G. Lambiase, Phys. Lett. B,  {\bf 715}, 1, (2012).
 

 




\end{thebibliography}
\end{document}